\documentclass{article}
\usepackage{setspace}
\usepackage{graphicx}
\usepackage[super]{natbib}
\usepackage{amssymb, latexsym}
\usepackage{amsmath} 
\usepackage{amsthm}
\usepackage{bm}
\usepackage[mathscr]{eucal}
\usepackage{enumerate}
\usepackage{multirow} 
\usepackage[center]{caption}
\usepackage{musicography}
\usepackage{float}

\usepackage{geometry}

\usepackage{authblk}
\usepackage{abstract}

\begin{document}
	
	\title{Impulse Pattern Formulation (IPF) Brain Model}
	
	\author{Rolf Bader}
	\affil{ Institute of Musicology\\ University of
		Hamburg\\ Neue Rabenstr. 13, 20354 Hamburg, Germany\\
	}
	\date{\today}
	
	\twocolumn
	[
	\begin{@twocolumnfalse}
		
		\maketitle
		\begin{abstract}
			A new brain model is introduced, based on the Impulse Pattern Formulation (IPF) already established for modeling and understanding musical instrument and rhythm perception and production. It assumes the brain works with impulses, neural bursts, ejected from an arbitrary reference point in the brain, arriving at other reflecting brain regions, and returning to the reference point delayed and damped. A plasticity model is suggested to adjust reflection strength in time. The model is systematically studied with 50 reflection points by varying the amount of excitatory vs. inhibitory neurons, the presence or absence of plasticity or external sensory input, and the strength of the input and plasticity in terms of system adaptation to an input or to the system itself. The Brain IPF shows adaptation to an external stimulus, which is stronger without plasticity, showing the active brain not being a simple passive \emph{tabula rasa}. A relation of 10-20\% of inhibitory vs. excitatory neurons, as found in the brain, shows a maximum adaptation to an external stimulus compared to all other relations, pointing to an optimum of this relation concerning adaptation. Although the model has no fixed timescale, when assuming strong brain periodicities only up to about 100 Hz, the reflection strength of the model is highest for delays of around 300 ms, corresponding to Event-Related Potential (ERP) timescales of brain potentials most often found roughly between 100 - 400 ms. The mean convergence times of the model correspond to short-time memory time scales with a mean of five seconds for converging IPFs. The Brain IPF is computationally very cheap, highly flexible, and with musical instruments already found to be of high predictive precision. Therefore, in future studies, the Brain IPF might be a model able to understand very large systems composed of an ensemble of brains as well as cultural artifacts and ecological entities. 
		\end{abstract}

\end{@twocolumnfalse}
]		
		
		\vspace{1cm}
		
		
		\renewcommand \thesection{\Roman{section}}
		\renewcommand \thesubsection{\Alph{subsection}} 
		\renewcommand \thesubsubsection{\arabic{subsubsection}} 
		
		\section{Introduction}
		
		Existing brain models vary in complexity, scaling, and mathematical modeling\cite{Kacprzyk2015}. On a low level, models solving differential equations of single neurons have been proposed like the Hodgkin-Huxley\cite{Hodgkin1952} or the Izhikevich model\cite{Izhikevich2007}, where, e.g., coincidence detection of spikes leaving the cochlear have been modelled\cite{Bader2018b}. Dynamical networks have also been proposed, modeling music large-scale forms using a FitzHugh-Nagumo (FHN) model\cite{Sawicki2022}, based on the empirical finding using EEG measurements of synchronization of brain parts with increasing musical tension\cite{Hartmann2014}. The transition of mechanical waves on the basilar membrane into neural spikes has shown phase synchronization between tone partials already on this low level\cite{Bader2015}, which is already able to reproduce timbre features like a bifurcating spectrum in a so-called surface tone of a cello\cite{Bader2018a}. These methods can be considered bottom-up models as they try to understand the brain by modeling coupled single neurons.
		
		Top-down models using global working principles of the brain have also been proposed. Friston assumes the brain to work according to a free-energy principle, in analogy to thermodynamics, as a change of equilibrium and surprises, which is also applied to hearing and music\cite{Friston2013}. Haken uses methods of Synergetics to calculate Gestalt perception\cite{Haken2008} analytically. Baars assumes the brain works as a global workspace, where brain parts synchronize and de-synchronize\cite{Baars2013}. The appearance and role of consciousness are also discussed within this framework. Such synchronization is also found with expectation\cite{Buhusi2005} and musical tension\cite{Hartmann2014}\cite{Sawicki2022}.
		
		Neural networks have also been proposed as connectionist\cite{Grossberg1976a}, \cite{Grossberg1976b} or as self-organizing\cite{Kohonen1995} networks. Musical applications in the connectionist sense concern music analysis\cite{Gjerdingen1990} or composition\cite{Briot2020}. They have often proven to be capable of analyzing and synthesizing music realistically. Still, their disadvantage is the impossibility of deriving the reason for successful modeling from the learned neural weights. Furthermore, they often need large databases to succeed. On the other hand, self-organizing maps (SOM) allow deriving reasons for the classification and sorting of input data in terms of music theory\cite{Leman1997}, defining ethnic groups\cite{Bader2021b}, or in musical acoustics\cite{Plath2021}. These methods are increasingly used in Computational Phonogram Archiving\cite{Bader2019} classifying, sorting, or analyzing big databases in ethnomusicology\cite{Blass2019}, streaming platforms, or all kinds of audio archives\cite{Blass2020}.
		
		Consciousness and conscious content are found to be spatiotemporal patterns as found with olfactory perception\cite{Kozma2016} or auditory perception\cite{Ohl2001}\cite{Ohl2016}. The brain shows phases of periodicity interrupted by chaotic transient phases within milliseconds. The periodic phases show repeating activity patterns of neural clusters forming structures in space and time. Different conscious content is thereby represented by different spatiotemporal patterns.
		
		Enlarging brain dynamics into society was proposed by Freeman as the idea of a society of brains\cite{Freeman2014} as a course of culture. Using the IPF and enlarging this framework by including cultural artifacts has been proposed in a Physical Culture Theory\cite{Bader2021a}.
		
		Within this line of reasoning, the present paper proposes a brain model based on the Impulse Pattern Formulation (IPF). It has first been proposed as a general method for understanding and modeling musical instruments \cite{Bader2013}. As a general method, it can compare musical instrument families, like plucked, blown, struck, or bowed instruments\cite{Linke2019a}\cite{Linke2019b}. The method assumes musical instruments to work with impulses, short energy bursts that are produced at a reference point, travel through the instrument, are reflected at one or several instrument parts, and return as a damped impulse to the reference point, which reacts to this impulse by sending a new impulse. To arrive at a most general model, the shape of the impulses is not considered and can be inserted into the resulting impulse pattern later. 
		
		As a result of multiple send-outs and reflections, an impulse pattern results. This pattern is represented by a system variable g which is updated at each iteration time and interpreted as a time interval or amplitude. So a constant value of g is a steady periodicity, corresponding to a constant pitch. A chaotic impulse pattern corresponds to a transient, an initial transient, or one of tone transition\cite{Linke2021a}. Bi-stable states represent multiphonics\cite{Linke2019b} known from wind instruments like saxophones or clarinets, where a player produces two or more pitches with the instrument by choosing complex fingerings or playing at minimum or maximum blowing pressure necessary for normal playing.
		
		The IPF always assumes an arbitrary reference point. Investigating the role of different parts of a classical guitar in the initial transient of a guitar tone, for each guitar part, like top plate, back plate, inclosed air or ribs, a unique IPF using these reference points results in a set of impulses patterns, one for each guitar part. Relating these IPF with the vibration of the parts calculated with a Finite-Difference Time Domain (FDTD) method solving the differential equations for plates and air on a guitar geometry shows a close correspondence between the IPF and the FDTD transients in terms of length and complexity\cite{Bader2013}.
		
		The IPF can predict the highly complex behavior of zither initial transients when varying the table the zither is placed on, a crucial part of zither sound production. Different strings acting over three zither feet onto different tables result in very different initial transient lengths. Linear measurement techniques cannot explain these initial transient lengths in terms of impedance or eigenmodes. The IPF, on the other hand, is very precisely modeling the system\cite{Linke2021b}.
		
		The minimum bow force needed to arrive at a regular sawtooth motion in violin or cello bowing has been underestimated with existing models. Again, the IPF comes very close to experimentally measured minimum bowing pressures much better than existing models\cite{Linke2022}. This is due to the self-organizing nature of the bowing process which is not reflected with previous linear models but met with the IPF.
		
		These reference points have also been proved to correspond with measurements of the Laotian wind instrument \emph{khaen}\cite{Hoover2001}, a set of free reeds attached to tubes. When playing with low pressure, the instrument sounds with the lowest eigenfrequency of the reed. From a certain blowing pressure threshold on, the instrument sounds with the frequency of the tube, which is the normal playing type. An IPF taking both reference points can understand the role and interaction of reed and tube and the reason for one of them forcing the other to go along its eigenfrequencies\cite{Linke2019c}.
		
		The IPF has also successfully been implemented in music psychology by modeling two musicians playing together, where one musician is changing the playing tempo, and the other musician reacts to this change. With only two reflection points modeled, the IPF reproduces behavioral data in such cases of tempo following, by finding that musicians need two beats from the past to adapt to a tempo change\cite{Linke2021a}.
		
		Taking the existing brain models into consideration, the IPF is proposed for two reasons. 
		
		First, as it has already proven to work with musical instruments very precisely by at the same time computationally very cheap, it has the capability to model large and extensive systems. This might be a brain; this might be a society of brains and cultural artifacts like musical instruments in a musical ensemble, this might also be cultural large-scale entities like societies or ethnic groups. The scale-free property of the method thereby allows to focus on tiny details or model large entities in their general behavior.
		
		Second, it does not differentiate between the kind of impulses but takes sound impulses in musical instruments and spike trains in the brain as their physical reality, electric fields, and interactions. This allows reconsidering the gap between psychology and physics simply by reducing both to electrical fields. As this holds even for the ecosystem as well as for matter in general, the IPF might have the capability to model large systems within one dynamical formulation.
		
		To arrive at such a large-scale model, the IPF must be systematically studied as a neural model. Therefore, the present paper applies the IPF in parameter space to examine its behavior and usability as a brain model.
		
		\section{Method}
		
		The IPF brain model is presented and tested using a parameter space discussed below. Next to the inclusion of the relation between inhibitory and excitatory neuron concentration, four cases are studied, where only the last, the action case, is realistic. Still, to arrive at an estimation of the realistic behaviour of the model with and without plasticity and with and without external input, all cases are equally discussed. Note that the input is performed in this paper as a separate IPF of a musical instrument. In the long run, the instrument and the brain are not to be taken as separate entities, and IPFs, including both are aimed for. Still, for the sake of understanding a Brain IPF working on its own, in this paper, they are still implemented as separate entities.
		
		\subsection{Brain Impulse Pattern formulation (IPF)}
		
		The brain is modeled using N = 50 neurons. Each neuron is a reflection point, returning impulses from a starting neuron, a viewpoint neuron. The system state of the viewpoint neuron is g, which represents a time period and amplitude strength. Each reflection neuron i has a damping $\alpha_i$. The IPF, then is
		
		\begin{equation}
			g_t = g_{t-1} - \frac{1}{N} \left|\sum_{i=2}^{N} P_i \ln \alpha_t^i\ g_{t-i}\right| + w_2\ g^I_t\ .
			\label{IPF_Brain}
		\end{equation}
		
		Here the viewpoint neuron is i=1. Therefore, the reflections come from neurons i=2,3,4,...N. P$_i$ are the polarizations of the neuron, where P$_i$ = 1 is an excitatory and P$_i$ = -1 is an inhibitory neuron. Note that the sum of all reflections is normalized using the amount of neurons N. The model is discrete with time steps t=0,1,2,3... Therefore, the earlier states of the viewpoint neuron, which this neuron has sent out to the other neurons, are returning after a delay in a damped and polarized form.
		
		Eq. \ref{IPF_Brain} takes the absolute value of the sum. In the case of negative values of $\alpha_t^i\ g_{t-i}$, the logarithm would be complex, and therefore g would become complex. As the cause of the logarithm in the IPF model is exponential damping, an exponential function with negative, real exponent\cite{Bader2013}\cite{Linke2019a}, a complex value of the exponent would also mean a harmonic oscillation, caused by the imaginary part of g, next to the exponential decay, caused by the real part of g. In other words, a negative $\alpha_t^i\ g_{t-i}$ would start a neural self-oscillation. Now, taking the absolute value of the sum means still taking the exponential damping, the real part of g, while also including the frequency of that oscillation, the imaginary part of g. As the system parameter g is a time interval, including also the imaginary part of g fits perfectly well into the interpretation of g. Indeed, when using the absolute value of the sum, negative values of g do not occur, and self-oscillation is turned off entirely in the present mode. 
		
		Taking the absolute value of the sum in Eq. \ref{IPF_Brain} also turns negative real parts of g into positive ones. Indeed, when wanting to get rid of the imaginary part of g, one could think of only using the real part of the sum. Still, this would lead to a negative g. But negative g are unphysical, as they would mean negative time intervals and negative amplitudes. Still, this is not too much of a problem, as $Re[\ln(-x)] = Re[\ln(x)]$, and therefore negative g would lead to the same system behavior. Negative g would also not lead to a positive exponent and a blow-up of the iteration, as the logarithm turns negative g into complex numbers, therefore, into oscillations, as discussed above. In any case, negative g are also not occurring when taking the more plausible absolute values of the sum as proposed in this paper. Still, future extensions of the Brain IPF could consider self-oscillations.
		
		\subsection{Plasticity Model}
		
		The plasticity of each neuron is calculated for each time step t. Plasticity means a change in the damping parameter $\alpha_i$, where each time step t then might have a different damping $\alpha_t^i$. Note that in the IPF reasoning, the damping is originally $1/\alpha$, which, for the sake of convenience, is skipped, now using $\alpha$ instead.
		
		For each time step, the new damping is calculated like
		
		\begin{equation}
			\alpha_t^i = \left|\alpha_{t-1}^i + w_1 \ln(1+(g_{t-i} - g_t))\right| \ .
			\label{IPF_plasticity}
		\end{equation}
		
		If the reflection point neuron t-i has the same value $g_{t-i}$  as the viewpoint neuron value $g_t$ the logarithm becomes zero, and no change in the damping $\alpha_{t^i}$ happens. If the reflection point neuron t-i has a larger value than the viewpoint neuron, the logarithm becomes larger than zero, and $\alpha_i$ increases. Otherwise, the logarithm assures a negative influence, and $\alpha_i$ decreases. The plasticity process is generally modeled using a constant $w_1$. Therefore, plasticity can be switched off in the model by using $w_1 = 0$. To examine different model behavior, $w_1$ will systematically be altered, shown below. Again, the absolute value of $\alpha$ is used, not allowing negative or complex values. This, again, does not change the model behavior due to the logarithms used. Still, positive values are more convenient. Indeed, negative arguments of the logarithm in the simulations shown below appear very rarely and are additionally suppressed by using the absolute value.
		
		\subsection{External Musical Instrument Input IPF}
		
		To examine the reaction of the Brain IPF to an external input of a musical instrument, a simple IPF is used, which models wind instruments or a piano in the case of normal playing conditions. The time development of the system parameter G$^{I}$
		
		\begin{equation}
			g^I_t = g^I_t - \ln \alpha^I\ g^I_{t-1} \ 
			\label{IPF_Instrument}
		\end{equation}
		
		is used with $g^I_0 = 1$ and $\alpha^I = 1.8$. This leads to a converged $g^I$ in time and a stable tone with constant pitch after a short initial transient phase of about 12 iterations. This musical instrument is fed into the Brain IPF as shown in Eq. \ref{IPF_Brain} with a coupling constant $w_2$. Again, for $w_2 = 0$, this input can be switched off.
		
		\subsection{Parameter Space}
		
		To systematically examine the behavior of the Brain IPF and its usability as a brain model for music, four cases, each with a certain parameter space, are examined:
		
		\begin{itemize}
			\item Static: Brain IPF \textbf{without plasticity} and \textbf{without instrument input}
			\item Plasticity: Brain IPF \textbf{with plasticity} but \textbf{without instrument input}
			\item Input: Brain IPF \textbf{without plasticity} but \textbf{with instrument input}
			\item Active: Brain IPF \textbf{with plasticity} and \textbf{with instrument input}
		\end{itemize}
		
		In the plasticity case the plasticity strength $w_1$ is varied as $0 \leq w_1 \leq 2$. In the input case the instrument input strength $w_2$ is varied like $0 \leq w_2 \leq 2$, both in steps of 0.1. In the active case, both $w_1$ and $w_2$ are varied accordingly. So the plasticity and input cases are special cases of the active case with $w_1 = 0$ or $w_2 = 0$ respectively.
		
		The initial values for the reflection strength $\alpha$ are randomly chosen for all neurons between $0 \leq \alpha_i \leq 1$ for i = 2,3,4...N, except for the viewpoint neuron, which has $\alpha_1 = 1$ and $P_1$ = 1. The initial choice of polarization, whether a neuron is excitatory or inhibitory, depends on the relation $C_{ex/in}$ between both neuron groups. Physiologically, 10-20\% of neurons are inhibitory\cite{Buzsaki2006}. Still, locally in certain brain regions, this relation might show larger deviations. To get an overview and understand the model behavior over the whole range of possible concentrations, relations $0 \leq C_{ex/in} \leq 1$ are varied, again in steps of 0.1 \%.
		
		The IPF might show considerably different behavior for different randomly chosen values of $\alpha$ and $g_0$, the initial value of the system parameter. To account for this, for each parameter combination of the parameter space, R=1000 cases were randomly chosen. Thereby, $0.5 \leq g_0 \leq 2$ was randomly determined. To follow the model’s output when making changes in the four cases or the parameter spaces, at first, 1000 sets of $\alpha$ with $g_0$ were produced. Then, for all cases and all parameter combinations in the parameter space, always these 1000 sets were used. This allows the following model behavior changes independent of the randomly chosen sets.
		
		\subsection{Detection of System Behavior}
		
		The study aims to discuss the adaptation of the brain to an external stimulus. Adaptation is defined as the convergence of the model. Convergence is achieved if
		
		\begin{equation}
			g_t = g_{t-1} \ ,
		\end{equation}
		
		where
		
		\begin{equation}
			\frac{1}{N} \left|\sum_{i=2}^{N} P_i \ln \alpha_t^i\ g_t\right| =  w_2\ g^I_t\  ,
			\label{Eq_Convergence}
		\end{equation}
		
		therefore, the input and the reflections cancel out.
		
		Note that in Eq. \ref{Eq_Convergence}, the system parameter g is that of $g_t$, omitting the delays at t-I, which means the convergence of g down to N delayed reflections. The input in the present study is a musical instrument IPF which converges after about 12 iterations. Therefore $w_2\ g^I_t = const$ thereafter. As the system parameters, g and $g^I_t$ are constant at convergence, condition Eq. \ref{Eq_Convergence} means a balance between g and the $\alpha_t^i$. As the reflection strength, $\alpha$, is subject to plasticity, a closed analytical solution of the system is no longer possible. Therefore, an iterative method is used below to estimate the system behavior.
		
		The output of a Brain IPF is a time series of the system parameter g and a time-dependent reflection strength $\alpha_t^i$. The cases without plasticity have constant $\alpha_t^i$ throughout. Still, the plasticity and the active cases also have a time dependency of $\alpha_t^i$.
		
		To account for all cases discussed above, the system parameter g is written as
		
		\begin{equation}
			g^{w_1,w_2,C_{ex/in},i}_{Case, t} \ ,
		\end{equation}
		
		Where Case is one of the four cases discussed above, $w_1$ and $w_2$ are the plasticity and instrument input strength, respectively, $C_{ex/in}$ the relation of inhibitory vs. excitatory neurons, i is the case of randomly chosen $\alpha$ and $g_0$, and t is the iteration time point. Only the active case uses both $w_1$ and $w_2$. With the plasticity and input cases, only one parameter is shown as an index, and in the static case, both indexes are omitted. This is easy to read as $C_{in/ex}$ is always given with a percent sign.
		
		The development of g can basically be of four types which might appear at the same time or adjacent:
		
		\begin{itemize}
			\item Converging: a constant value of g is reached over time
			\item Periodic: a periodicity appears, more or less stable
			\item Complex: noise-like time series
			\item Amplitude jumps: sudden amplitude jumps appear, with overall periodic or complex behavior
		\end{itemize}
	
	\begin{figure}
		\centering
		\includegraphics[width=.83\linewidth]{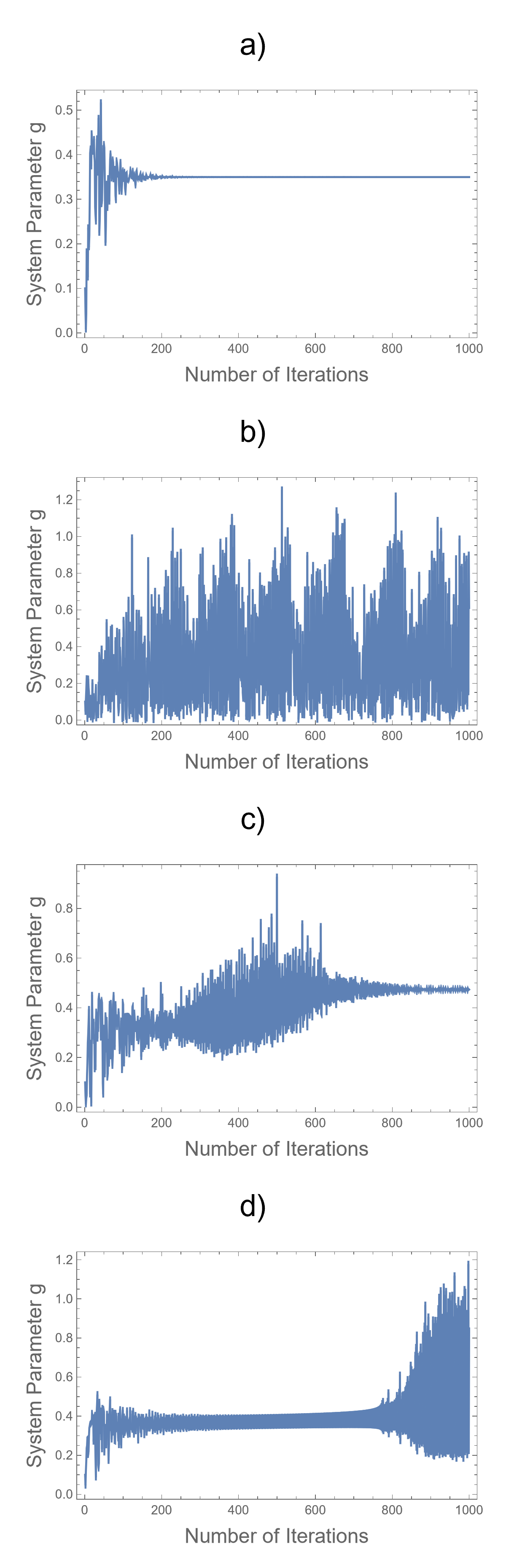}
		\caption{Examples of time series g of the Brain IPF showing a) convergence, b) divergence with periodicity with complexity, c) convergence with complexity and sudden amplitude jumps, and d) divergence after a time of periodicity.}
		\label{fig:gexamples}
	\end{figure}
	
		Fig. \ref{fig:gexamples} shows four examples for time series of the system parameter g for random $g_0$, $\alpha$, and $C_{in/ex}$. In Fig. \ref{fig:gexamples} a) a converged case is shown. In b) a non-converged case with a periodicity and much complexity is displayed. The c) plot shows a case of slow convergence with sudden amplitude jumps, periodicity, and complexity. Finally, in d) a case of periodicity appearing after a complex initial transient is shown which suddenly diverges. Clearly, a precise association to one of the phenomena appearing is only possible with the convergence case which is used in the present paper to address adaptation. 
		
		Of course, these types can be differentiated even more and are not perfectly defined. Still, this study’s purpose is to adapt the brain to an external musical instrument input. As we know that this input is periodic after an initial transient, which as IPF, means a convergent system parameter g, in this paper, we concentrate on the detection of a converging time series of brain dynamics. Therefore, we only detect if an IPF is converging or not.
		
		Convergence is detected by calculating the standard deviation of g over the last 200 values of g. All IPFs in this paper are calculated with 50 reflection points. As the system parameter g also is a time interval, the length of an IPF with T iterations depends on the resulting g. So e.g., if a musical tone with a fundamental frequency of 100 Hz is used as input, the model considers immediate brain responses within 0.5 seconds. Of course, due to impulses traveling through all reflection points, the 'echo' of a system state g or an external input in the system is much larger. The choice of 50 reflection points has been determined beforehand by varying this amount from one up to 1000 reflection points. There a convergence of the system behavior up to 50 reflections appeared. Still, a systematic evaluation of the amount has yet to be performed which might be necessary for future work.
		
		The  neocortex frequencies are measured with reasonable strength and associated with cognitive content only up to about 100 Hz with the highest gamma-band present from about 30-70 Hz. In the auditory pathway frequencies, up to 4 kHz are processed from the cochlear on\cite{Bader2015}, but beyond the A1 region of the auditory cortex, frequencies are much lower. Still, in future studies of the auditory pathway, the amount of reflection points might need to be increased to cover higher frequencies.
		
		Convergence of IPFs is calculated as standard deviations over all R cases at iterations 800 to 1000 like
		
		\begin{equation}
			S_{Case}^{w_1, w_2, C_{in/ex}} = \frac{1}{R} \sum_{j=1}^R std(g^{w_1,w_2,C_{in/ex},j}_{Case, 800-1000}) \ ,
		\end{equation}
		
		For j=1,2,3,...R, where std is the standard deviation. When averaging over $w_1$ and $w_2$ in those cases these parameters are varied we arrive at $\overline{S}_{Case}^{C_{in/ex}}$.
		
		The mean values for the system parameter themselves are respectively
		
		\begin{equation}
			M_{Case}^{w_1, w_2, C_{in/ex}} = \frac{1}{R} \sum_{j=1}^R mean(g^{w_1,w_2,C_{in/ex},j}_{Case, 800-1000}) \ ,
		\end{equation}
		
		and $\overline{M}_{Case}^{C_{in/ex}}$.
		
		In the model, 50 reflection points are defined. To arrive at an overview of the variations of the reflection strength $\alpha$ when adding plasticity to the model, a mean is defined as
		
		\begin{equation}
			\overline{\alpha}^{w_1, w_2, C_{in/ex}}_{Case} = mean(\alpha_{t=1000, Case}^{i, w_1, w_2,  C_{in/ex}}) \ ,
		\end{equation}
		
		where $\alpha_{t=1000, Case}^{i, w_1, w_2,  C_{in/ex}}$ are the $\alpha$ values at the latest iteration time point t = 1000 for all cases and plasticity and input strength variations $w_1$ and $w_2$ respectively, where present in the cases. Again, the variable $1 \leq i \leq 50$ represents the $\alpha$ at the respective reflection point in the model, as discussed above. So $\overline{\alpha}^{w_1, w_2, C_{in/ex}}_{Case}$ is integrating over all reflection points i.
		
		It is also interesting to compare earlier and later reflection point strength. This is computed for the plasticity case, where overall $w_1$ is integrated like
		
		\begin{equation}
			\overline{\alpha}^{i, C_{in/ex}}_{Plasticity} = mean(\alpha_{t=1000, Case}^{i, w_1, w_2,  C_{in/ex}}) \ .
		\end{equation}
		
		Another measure is the time at which convergence has been established. This is again calculated as a mean over R=1000 cases of randomly chosen $\alpha$ and initial system parameter $g_0$ values like
		
		\begin{equation}
			T_{Case}^{w_1, w_2, C_{in/ex}} =  \sum_{j=1}^{R conv} mean\left(P^{w_1,w_2,C_{in/ex},j}_{Case}\right) \ ,
		\end{equation}
		
		where $P^{w_1,w_2,C_{in/ex},j}_{Case}$ are the time points the time series g has converged. The convergence time point is chosen to be the first time point where $mean\left(g^{w_1,w_2,C_{in/ex},j}_{Case, i\ to\ i+10} \right)< 0.01$. So the mean is taken over ten adjacent values of g. The threshold of 0.01 is chosen concerning the overall range of appearing values of g. The input sound, also an IPF, reaches this convergence criteria after 12 iterations. The mean for convergence is not taken over all R=1000 cases but only over those meeting the convergence criteria. All others are considered not to converge. 
		
		Finally, the minima convergence time points are chosen like
		
		\begin{equation}
			\underline{T}_{Case}^{w_1, w_2, C_{in/ex}} = min\left(P_{Case}^{w_1, w_2, C_{in/ex},j}\right)\ .
		\end{equation}

		\section{Results}
		
		For all R=1000 sets of $\alpha^i$ and $g_0$, all four cases with all parameter variations $w_1$ and $w_2$ and all inhibitory/excitatory neuron concentrations $C_{ex/in}$ IPFs are calculated and convergences $S_{Case}^{w_1, w_2, C_{in/ex}}$, $M_{Case}^{w_1, w_2, C_{in/ex}}$, $\overline{\alpha}^{w_1, w_2, C_{in/ex}}_{Case}$, $\overline{\alpha}^{i, C_{in/ex}}_{Plasticity}$, $T_{Case}^{w_1, w_2, C_{in/ex}}$, and $\underline{T}_{Case}^{w_1, w_2, C_{in/ex}}$ and the means $\overline{S}_{Case}^{C_{in/ex}}$ and $\overline{M}_{Case}^{C_{in/ex}}$ are determined. To get an overview, at first, the variations of $C_{in/ex}$ are discussed concerning overall convergence. Then, system behavior with varying plasticity and input strengths are discussed followed by a discussion of the shape of adapted reflection point strength. Finally, convergence time is discussed in terms of short-time memory effects.
		
		\subsection{Overall Model Convergence Behavior}
		
		A typical relation of inhibitory vs. excitatory neurons in the brain is about 10-20\%\cite{Buzsaki2006}. Still, for an overview, it is of interest how the model behaves with all possible concentrations. Indeed, in some cases, when modeling only a part of the brain, different relations might be present.
		
\begin{figure}
			\centering
			\includegraphics[width=1.05\linewidth]{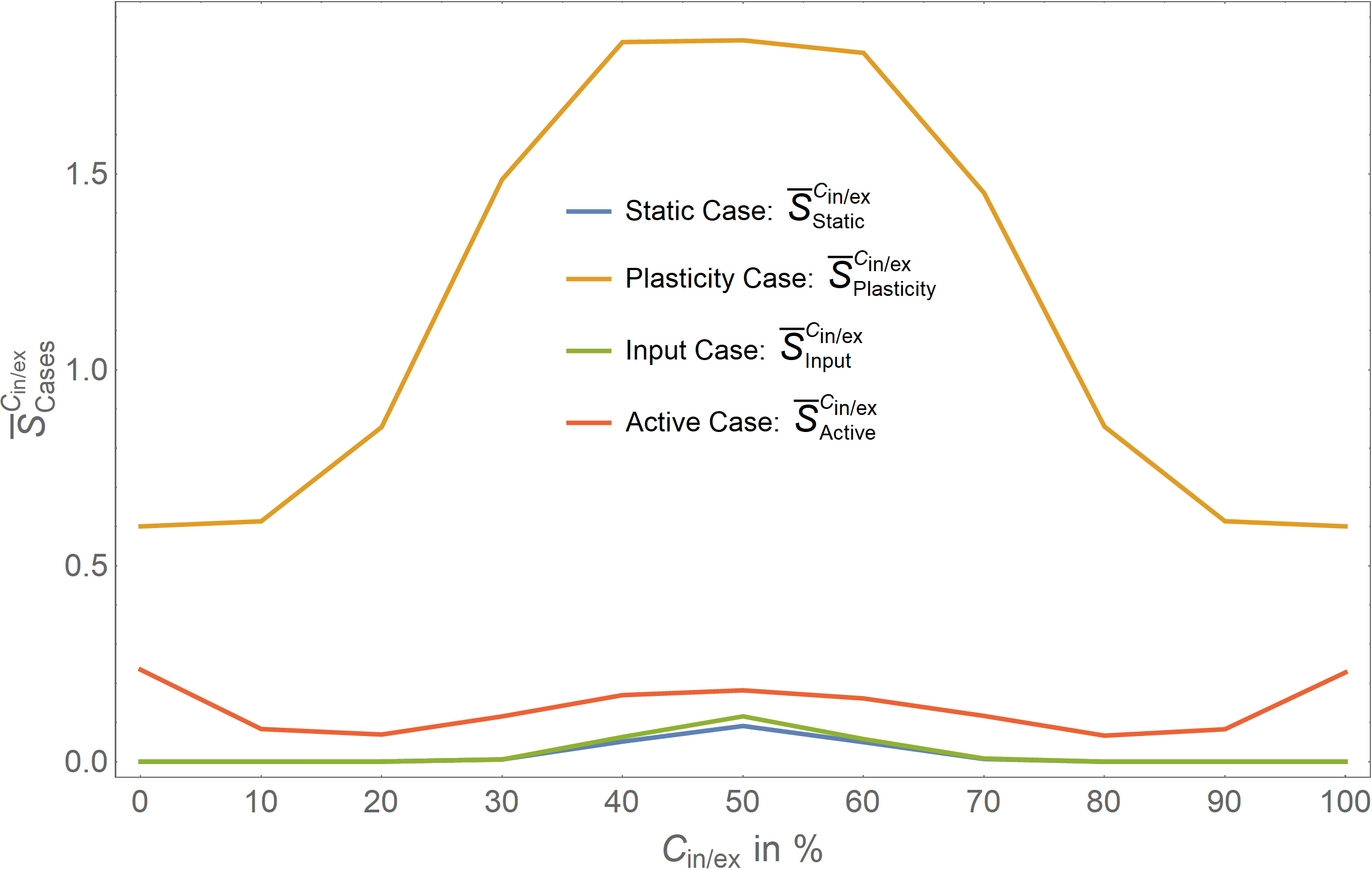}
			\includegraphics[width=1.05\linewidth]{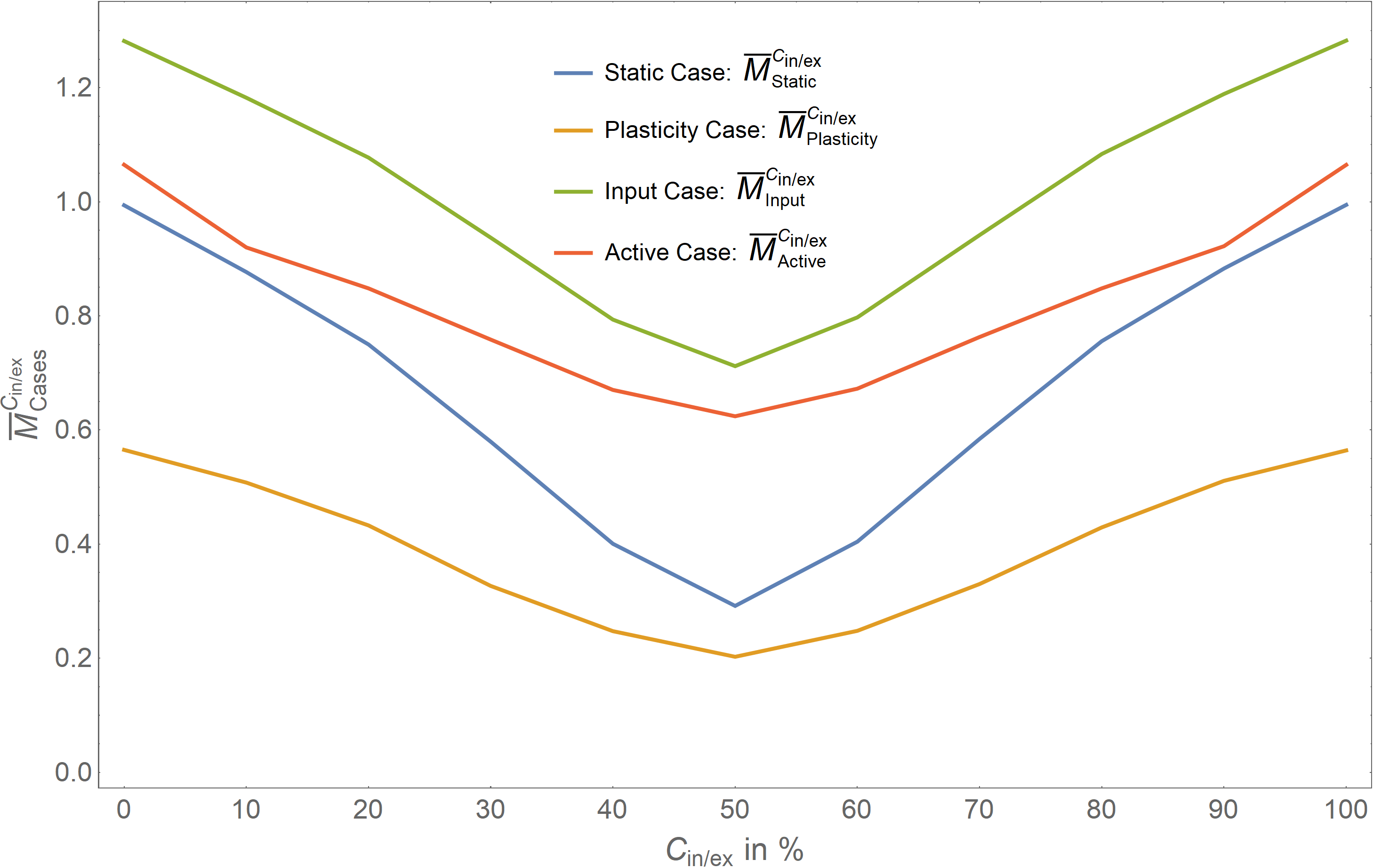}
			\caption{Standard deviation $\overline{S}_{Case}^{C_{in/ex}}$ (top) and mean $\overline{M}_{Case}^{C_{in/ex}}$ (bottom) of the system parameter g for iterations 800-1000, averaged over all R=1000 cases of randomly chosen $\alpha$ and $g_0$ and variations of coupling strength $w_1$ and $w_2$ for the static, plasticity, input, and active cases. Therefore the plots give a rough overview of the convergence of the model. When varying the amount of inhibitory vs. excitatory neurons in the model, for convergence, a peak at 50\%, and for mean, a minimum appears in all cases. Still, for the physiologically realistic active case, this peak is the smallest and overall smooth behavior is found.}
			\label{fig:meanstdalle}
		\end{figure}

		Fig. \ref{fig:meanstdalle} shows at the top plot $\overline{S}_{Case}^{w_1, w_2, C_{in/ex}} $, the averaged standard deviations of the system parameter g over iterations 800-1000. This parameter shows the overall convergence behavior of the model. As a first rough overview, the averaged convergence is taken over all R=1000 cases of $\alpha$ and $g_0$ and over the variations of $w_1$ and $w_2$ when present in the respective cases.
		
		In all cases, a peak at a 50\% relation of $C_{in/ex}$ is found. The static case has the strongest convergence, nearly the same when giving an input to the model. This is expected as the input itself has fast convergence. When adding plasticity to the model, the convergence is decreased tremendously. The brain model's activity is counteracting the static case's convergence behavior. The convergence is again much stronger when adding the input, but without arriving at the original convergence strength of the static case. 
		
	An opposite behavior is found for the mean system parameter $\overline{M}_{Case}^{w_1, w_2, C_{in/ex}}$ shown in the bottom plot of Fig. \ref{fig:meanstdalle}. There, a minimum is found in all cases at $C_{in/ex}$ = 50 \%. This points to a balance of inhibitory and excitatory neurons at 50\%, leading to low values of the system parameter. 
	
	Again, the plasticity case is the  most extreme, but the values are lowest here. The input case has the largest values as expected, as the input gives additional energy into the system. The active case is again medium and, therefore, again balancing between the plasticity and the input case. 
	
	Therefore, the model behaves as expected from a neural network in general.
		
		\subsection{Convergence when varying Plasticity\newline Strength}
		
		In the plasticity case, only the plasticity strength $w_1$ is varied. Fig. \ref{fig:stdplasticity} shows the convergence $S_{plasticity}^{w_1, w_2, C_{in/ex}}$ for 0\% $\leq C_{in/ex} \leq$ 50\% variations of the relation between inhibitory and excitatory neuron concentration. For realistic concentrations of 10 - 20\%, a strong decrease of convergence is found with low plasticity strength $w_1$. Then, with $w_1 > 0.3$, convergence is strong again. With higher concentrations, starting from 30\%, this peak is much eased and appears with higher $w_1$. The mean system parameter $M_{Plasticity}^{w_1, C_{in/ex}}$ is decreasing with increasing convergence.
		
		This behavior points to an instability caused by the interplay of differently delayed neurons. 
		
		\begin{figure}
			\centering
			\includegraphics[width=1\linewidth]{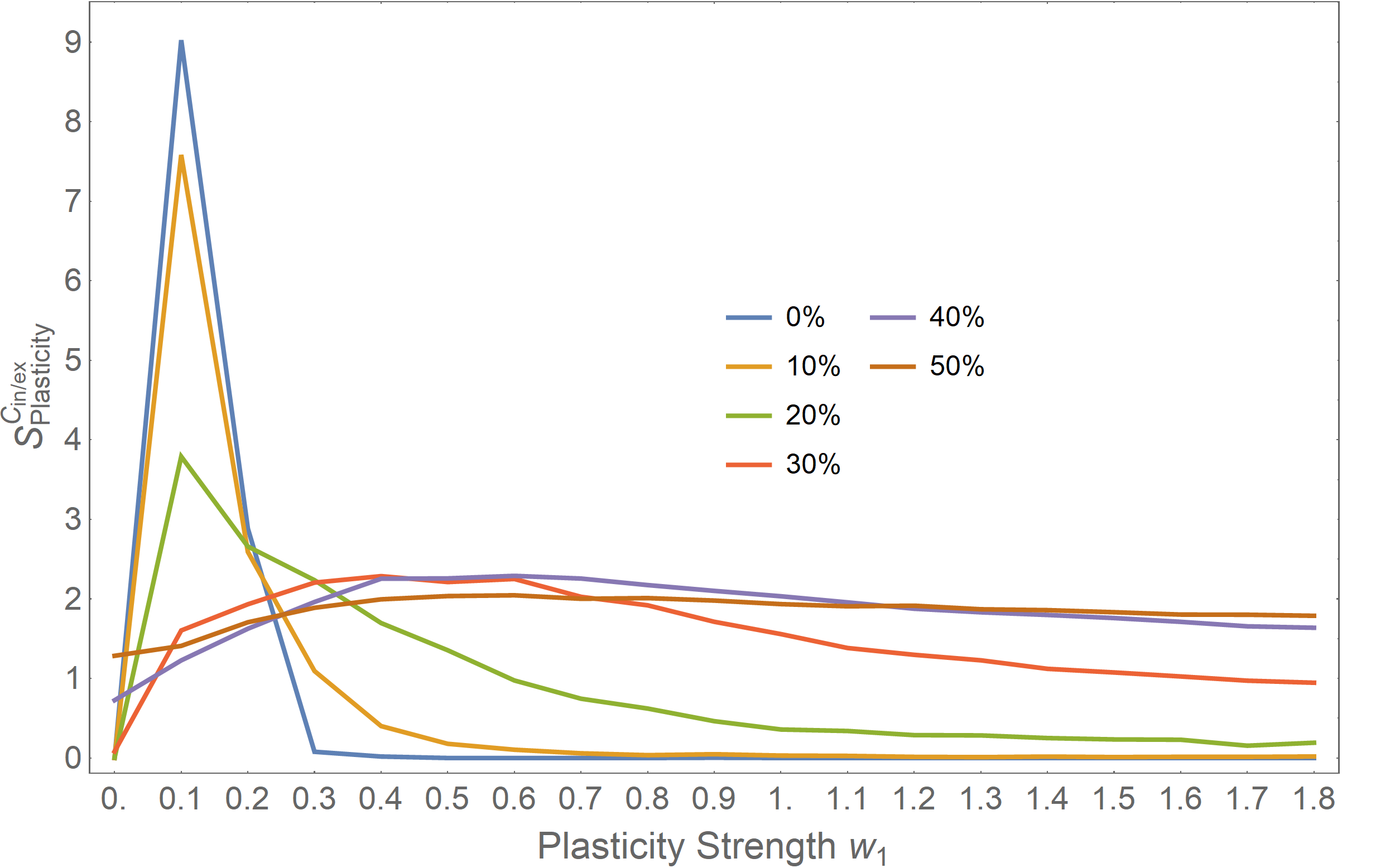}
			\includegraphics[width=1\linewidth]{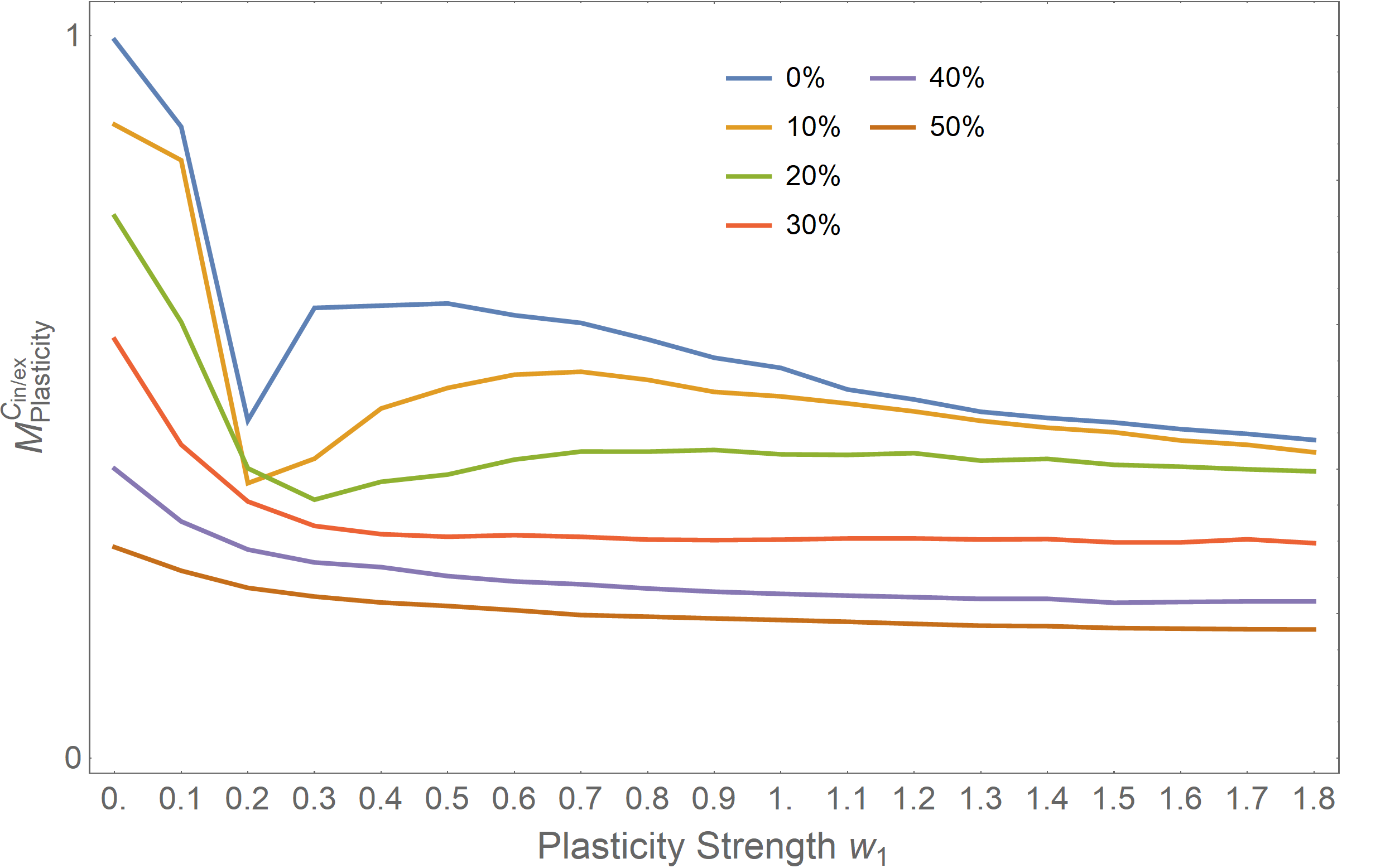}
			\caption{Model convergence when varying plasticity strength $w_1$ in the plasticity case (top plot) and the mean system variable g in this case (bottom plot). The convergence in the plasticity case is less converging for weak plasticity strength $w_1$ and realistic $C_{in/ex}$ concentrations. Accordingly, $M_{Plasticity}^{w_1, C_{in/ex}}$, the mean of the system variable g is decreasing with increasing convergence.}
			\label{fig:stdplasticity}
		\end{figure}
		
		\subsection{Convergence when varying Input\newline Strength}
		
		In the input case, the situation is more apparent. Increasing $w_2$ leads to stronger convergence. As found before for the input case, $C_{in/ex}$ concentrations around 50\% converge less but are also eased with higher $w_2$. Still, the mean system parameter $M_{Input}^{w_2, C_{in/ex}}$ is increasing with decreasing convergence. This is the opposite behavior compared to the plasticity case, where the system parameter decreased with increasing convergence. The additional energy of the input to the system can explain this.
		
		\subsection{Neuron Reflection Point Strength\newline Delay-Dependency}
		
		When plasticity is added to the model, in the plasticity and the active case, the convergence of $\alpha$ is of interest. At t = 1000 iterations, so at the end of the chosen simulation length, the alpha randomly selected at the beginning will have converged in all converging cases of the system parameter g.
		
		\begin{figure}
			\centering
			\includegraphics[width=1\linewidth]{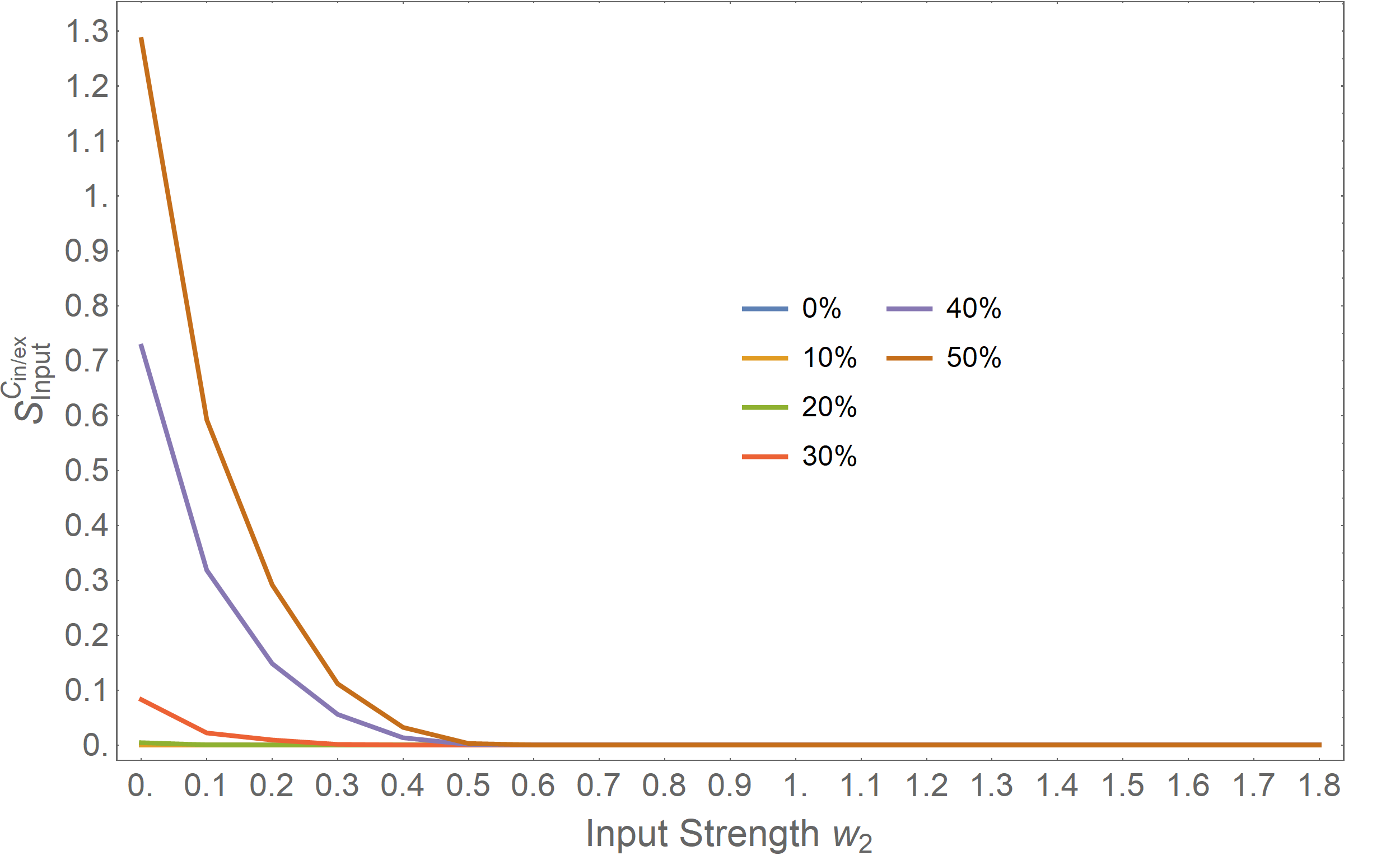}
			\includegraphics[width=1\linewidth]{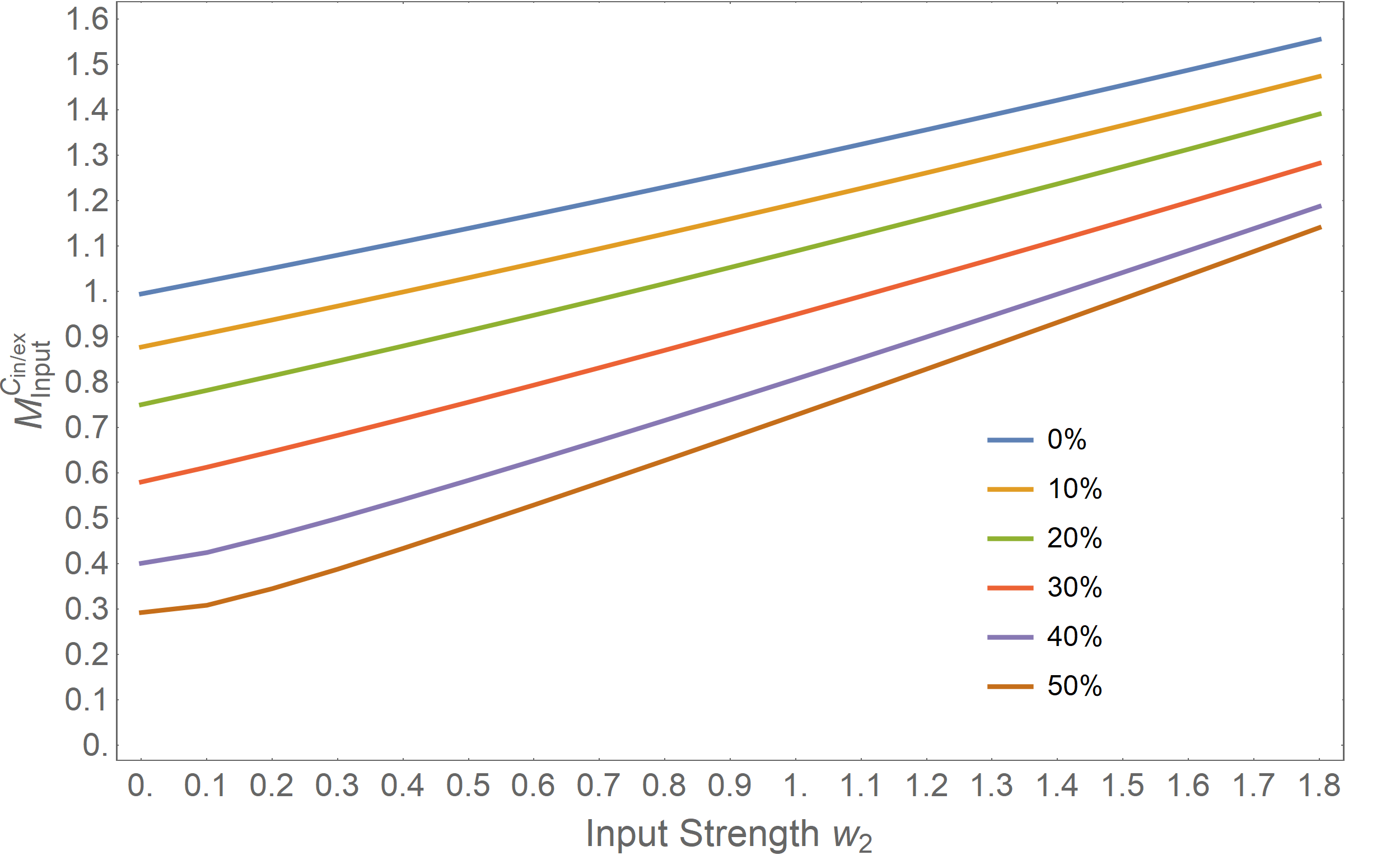}
			\caption{Model convergence when varying the input strength $w_2$ in the input case. The input case behaves as expected by increasing convergence with higher $w_2$. Contrary, $M_{Input}^{w_2, C_{in/ex}}$ is increasing with higher input strength $w_2$.}
			\label{fig:stdplasticity}
		\end{figure}
		
\begin{figure}
			\centering
			\includegraphics[width=1\linewidth]{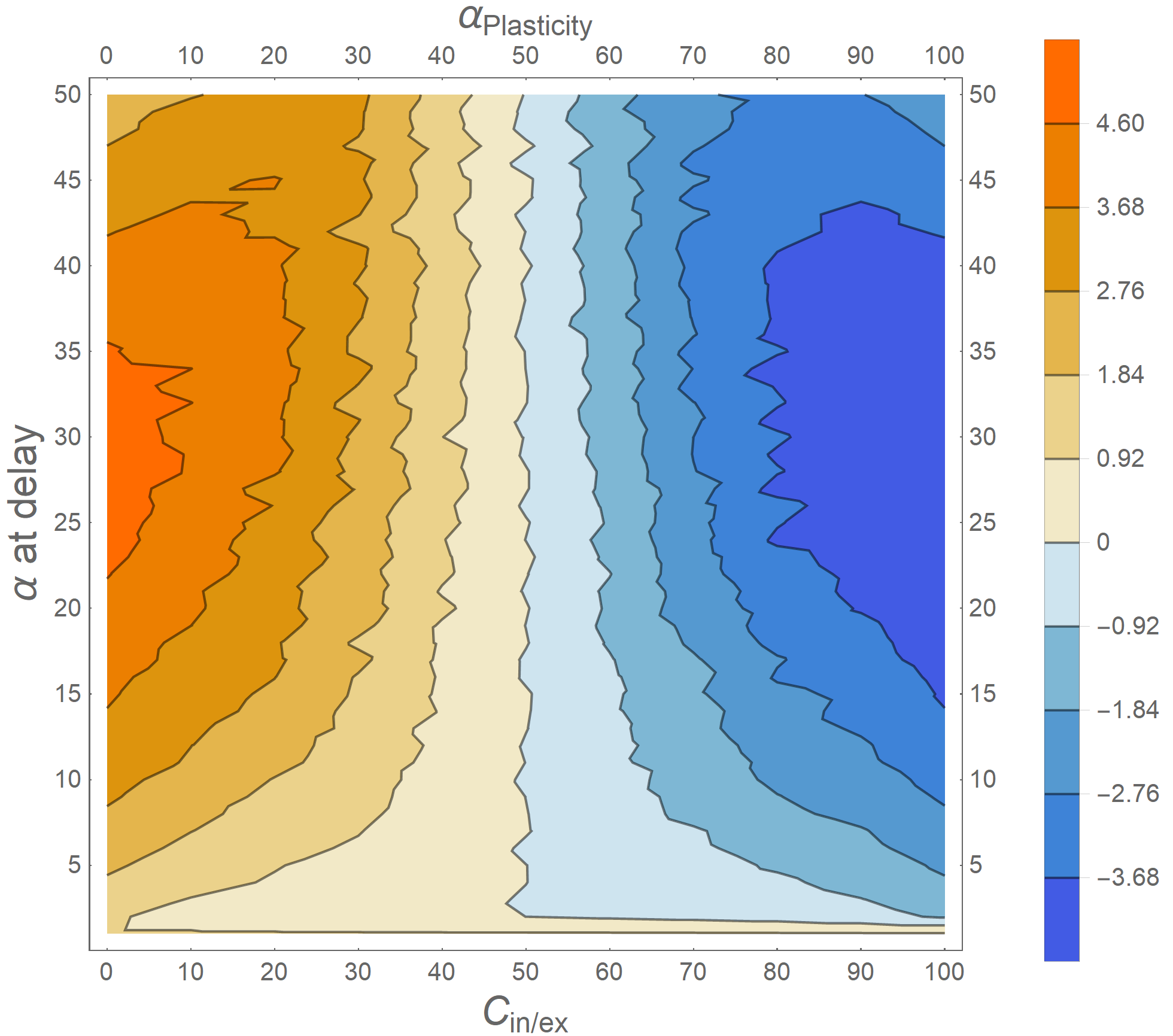}
			\caption{Mean of $\alpha$ at the 50 reflection points at the latest time point t = 1000 for the plasticity case and all $C_{in/ex}$. For a greater amount of excitatory neurons on the plot’s left side, the values are positive and vice versa. A peak of reflection strength is around reflection point 30 in all cases.}
			\label{fig:alphaplasticity}
		\end{figure}

		Fig. \ref{fig:alphaplasticity} shows $\overline{\alpha}^{i, C_{in/ex}}_{Plasticity}$, the integration over all plasticity strength $w_1$. As $C_{in/ex}$ determines the amount of positive and negative reflection points, the left side of the plot is positive while the right side is negative, as expected. Still, the reflection point strength is strongest around reflection point 30 in all cases. The uneven contour lines point to a large variation of this behavior. Still, overall in this model, reflections around reflection point 30 has the largest impact on the system behavior when the system adapts to itself without any input.
		
		\subsection{Neuron Reflection Point Strength with changing Plasticity and Input Strength}
		
\begin{figure}
			\centering
			\includegraphics[width=1\linewidth]{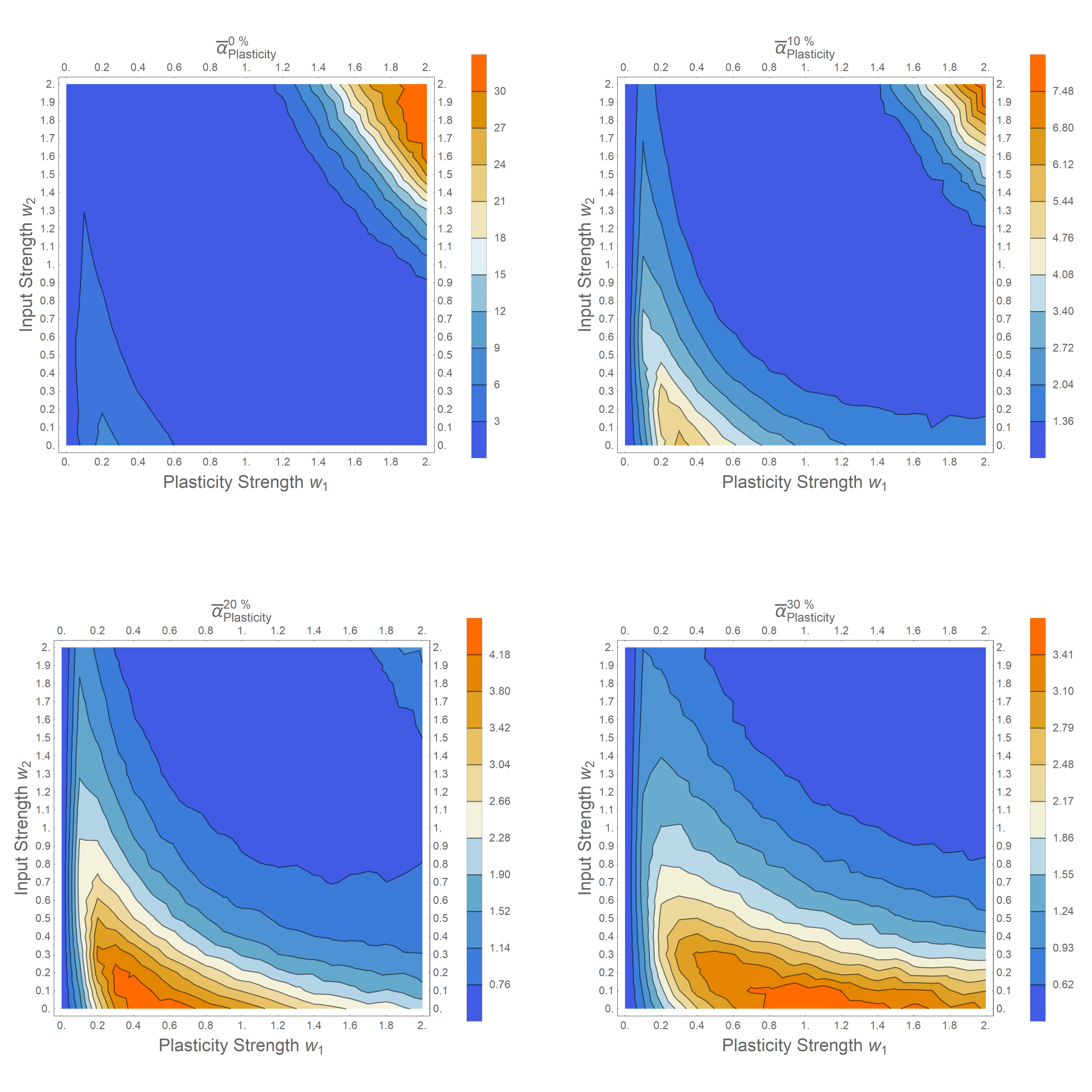}
			\caption{Mean neuron reflection point strength when varying plasticity and input strength $w_1$ and $w_2$ respectively for different inhibitory vs. excitatory neuron concentrations $C_{in/ex}$.}
			\label{fig:alphaw1w2}
		\end{figure}

		The adaptation of the reflection point strength $\overline{\alpha}^{w_1, w_2, C_{in/ex}}_{Active}$ when varying both plasticity and input strength $w_1$ and $w_2$ respectively is shown in Fig. \ref{fig:alphaw1w2} for four inhibitory vs. excitatory neuron concentrations $C_{in/ex}$. These are integrated over all reflection points.
		
		The range of values decreases with increasing $C_{in/ex}$ from over 30 down to about 3.5. The large values in the 0\% plot on the top left are mainly found with large $w_1$ and $w_2$, while for small plasticity and input strength values around 3-9 are found. These values decrease with increasing concentration. Also, the peak of $w_1$ with low $w_2$ is changing from around 0.2 at 0\% to about 1.1 at 30\%. This peaking behavior is also found with $w_2$, although not that strong. In the 30\% case and constant $w_1$, $\overline{\alpha}^{w_1, w_2, 30\%}_{Active}$ in- and decreases with increasing input strength $w_2$. 
		
		This is pointing to a more realistic model behavior with realistic $C_{in/ex}$ of 20 - 30\%. 
		
		\subsection{Convergence Time}
		
\begin{figure}
			\centering
			\includegraphics[width=1\linewidth]{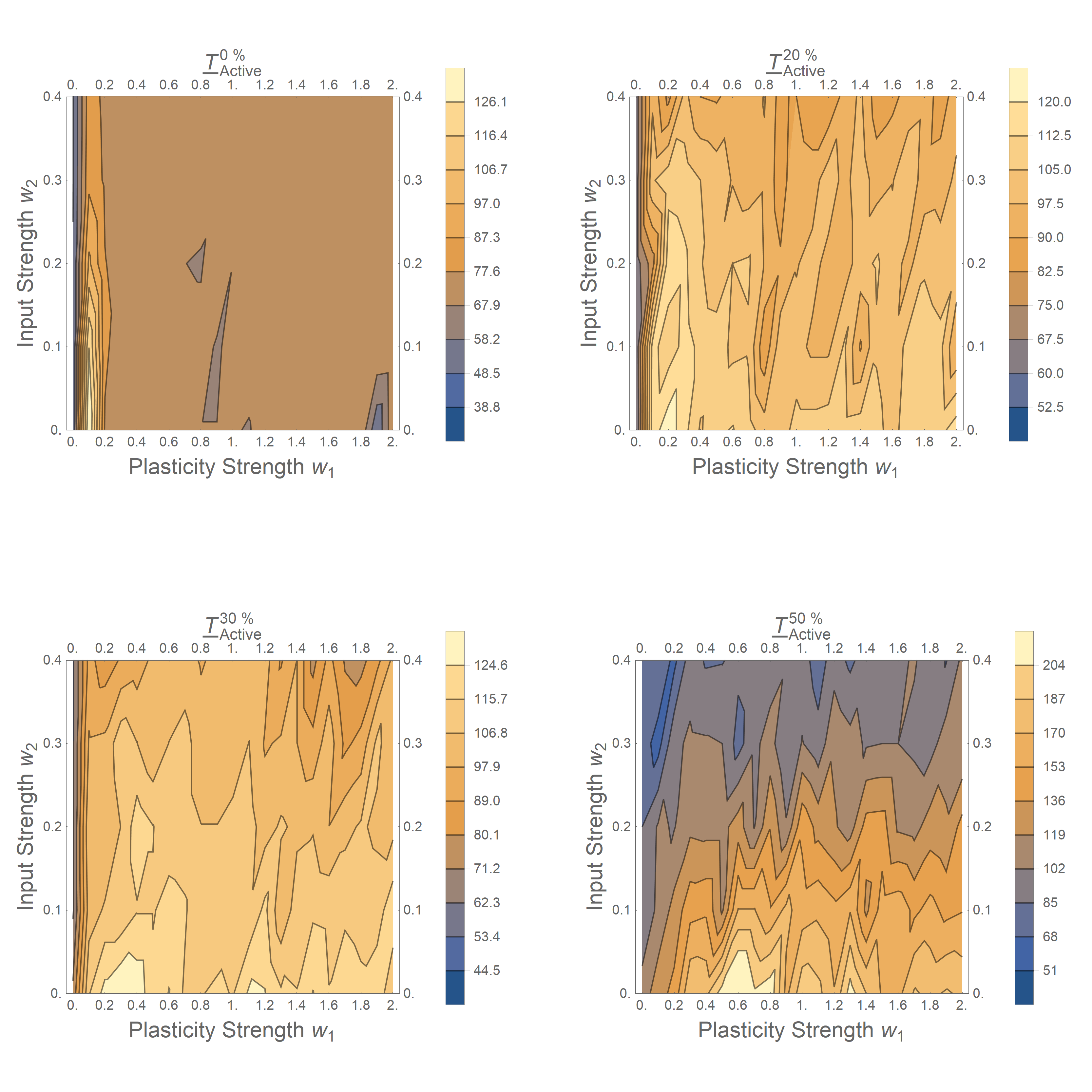}
			\caption{Minimum convergence time $\underline{T}_{Active}^{w_1, w_2, C_{in/ex}}$ of the active case when varying $w_1$ and $w_2$ in number of iterations. Taking a fundamental frequency of the input sound of 100 Hz, 100 iterations correspond to 1 second. Therefore, convergence times fit within short-time memory time.}
			\label{fig:stdconvtimemin-active}
		\end{figure}

		The convergence time has been defined as a mean and a minimum time. It is discussed here for the active case, as the input and plasticity cases are only special cases of the active case with $w_1 = 0$ and $w_2 = 0$, respectively, and therefore included in the discussion. The overall behavior of the mean convergence time is similar to that of the plasticity $\overline{\alpha}^{w_1, w_2, C_{in/ex}}_{Active}$ shown in Fig. \ref{fig:alphaw1w2} and therefore omitted here. The minimum convergence times  $\underline{T}_{Active}^{w_1, w_2, C_{in/ex}}$ are shown in Fig. \ref{fig:stdconvtimemin-active}. Note that the mean convergence times are calculated from only those R=1000 cases where between iterations 800-1000 convergence has appeared. All other cases are considered non-converging and excluded here. Therefore, discussing a maximum convergence time is not reasonable.
		
		The mean convergence time $T_{Active}^{w_1, w_2, C_{in/ex}}$ covers a range from about 100 to 500 iterations. Taking 100 Hz as the fundamental frequency of the input sound and an the iteration time interval of 10 ms per iteration of the system parameter g, 500 iterations corresponds to five seconds. This corresponds to short-time memory time. Although much smaller convergence times exist, the model is reasonably within typical brain behavior. Therefore, the Brain IPF explains short-time memory length using the physiological threshold of strong neural activity only up to about 100 Hz, as found in EEG signals.
		
		Taking a look at the minimum time point $\underline{T}_{Active}^{w_1, w_2, C_{in/ex}}$ as shown in Fig. \ref{fig:stdconvtimemin-active}, for $C_{in/ex}$ = 20 \% and 30 \%, the distribution is different from the mean cases, where a convergence of about 100 iterations, so about 1 second is found nearly all over variations of $w_1$ and $w_2$. Exceptions are only found for $w_1 = 0$. This is different for the 0 \% and 50 \% cases. For 0 \%, small $w_1$ lead to an increase of the minimum time, while in the 50 \% case, an increase of the input strength $w_2$ leads to a decrease of this minimum time. So in the realistic cases of 20 \% and 30 \% cases of inhibitory vs. excitatory neuron concentration, the minimum convergence possible is basically independent of plasticity and input strength. 
		
		\section{Conclusion}
		
		Viewing the brain as a system of spike bursts sent out by an arbitrary reference point to multiple other brain regions, which return the bursts in a damped manner, as implemented in the IPF Brain model discussed in this paper, leads to reasonable behavior. Converged states with a periodicity are possible, next to complex and chaotic states. With no sensory input, the case with plasticity tremendously increases the amount of complex brain activity while reducing the overall system energy compared to the case without plasticity. With sensory input added, the system strongly adapts to this input while simultaneously increasing overall system energy. Still, sensory input with plasticity means more complexity than sensory input without plasticity. Therefore, plasticity leads to less adaptation and more behavior intrinsic to the brain. This corresponds to an active, creative adaptation of the brain to an external stimulus rather than a passive \emph{tabula rasa}. 
		
		The amount of inhibitory vs. excitatory neurons in the brain is between 10 - 20\%. Still, in single brain units, the amount might be different. Here again, the active case of plasticity adapting to an external sensory input differs from all other cases. While the amount of complexity shows a maximum at $C_{in/ex}$ = 50\%, the active case has a minimum of complexity for this region of 10 - 20\%. The Brain IPF, therefore, finds this physiological finding in a minimization of brain complexity when adapting to sensory input.
		
		This preference for $C_{in/ex}$ = 10 - 20\% also appears in the plasticity case when varying the plasticity strength $w_1$. For 0\% $\leq C_{in/ex} \leq$ 20\% complexity decreases strongly for $w_1 <~ 0.3$, although for very low $w_1$ in this region complexity is much stronger compared to higher $C_{in/ex}$. As plasticity strength might vary for different brain regions, this behavior means a effective variability of brain behavior by varying $w_1$, which might also appear temporally, a variation beyond the scope of this paper.
		
		The brain regions' reflection strengths $\alpha$ are systematically distributed after convergence with a maximum strength at about reflection point 30, increasing to this maximum from the reference point on and decreasing towards reflection point 50. This is strongest for low $C_{in/ex}$. When using a sensory input with a periodicity of 100 Hz, a maximum of 30 reflection points corresponds to a maximum reflection strength around 300 ms. This is a timescale known from Event-Related Potentials (ERPs), where a reaction of a stimulus is found to be present in time intervals of roughly 100-400 ms, as e.g., found with Mismatch Negativity (MMN) experiments of musical syntax found in the Broca region\cite{Koelsch2011}, binaural Interaural Cross-Correlation (IACC) spatial perception neural mastoid response\cite{Lueddemann2007}, or used in understanding music perception of cochlear implant patients\cite{Sharp2018}. 
		
		Considering the convergence time points, the model meets the short-time memory time-span of up to about five seconds, often associated with an echoic memory\cite{Snyder2000}. A minimum with a ,reasonable input and plasticity strength of about one second is found. All this appear when taking 100 Hz as the fundamental frequency of the input sound and therefore meet the physiological finding of strong periodicities in the cortex only up to about 100 Hz. Thus, the model finds short-time memory length to come into place due to this 100 Hz maximum and the basic idea of the brain acting as an interchange of damped and delayed reflected impulses.
		
		Other time scales are known and important in the brain and in auditory perception. A periodicity of 50 Hz of oscillating neurons was found to be the basis of perception and cognition\cite{Buzsaki2006}\cite{Sawicki2022}\cite{Bader2021a}. This frequency is associated with the gamma-band which also convers higher frequencies up to about 100 Hz. Also, 50 Hz is half the frequency used in this model. Neurons firing at 50 Hz on their own and in asynchrony with other neurons also oscillating at 50 Hz will lead to a period-doubling of the overall oscillation leading again to 100 Hz. Therefore, the findings presented in this paper support a periodicity of 50 Hz more than arguing against it.
		
		The model is that of a neocortex, as it is referenced to EEG signals taken from the neocortex. Of course, adapting to an external sound input while perceiving a musical pitch is much faster and within the range of milliseconds. The second integration time of auditory and visual perception is around 50 ms, where successive events with shorter intervals are no longer perceived as separate. In terms of auditory perception, this leads to the lowest frequency of about 20 Hz, where rhythm perception ends, and pitch perception starts. Although pitch perception is not understood in terms of a general agreement until now, many findings point to a subcortical perception, including spaciousness or timbre perception\cite{Bader2021a}. Discussing this is beyond the scope of this paper. So a Brain IPF of the auditory pathway would need to use a much shorter time interval compared to the 100 Hz used in the present model of the neocortex. 
		
		The Brain IPF model, therefore, behaves reasonable within existing findings of brain activity. Its advantage compared to existing models is its very low computational cost and flexibility in terms of variations of plasticity strength, sensory input, or time scales. Another major advantage is a formulation already found working with musical instruments, so with cultural artifacts. This opens the possibility to model ensembles of brains and musical instruments or in the future even larger entities like societies or ecosystems within one model, which is computationally fast, highly flexible, and very precise in terms of system predictions, as already found with musical instruments. 
		
\section{Acknowledgement}
We thank Robert Mores and Simon Linke for their valuable comments on the paper.

\end{document}